\documentstyle{article}
\begin{document}

\author{Fabi\'an H. Gaioli,\footnote{E-mail: gaioli@iafe.uba.ar} 
Edgardo T. Garcia Alvarez, and Mario A. Castagnino \\
%EndAName
{\it Instituto de Astronom\'\i a y F\'\i sica del Espacio, }\\
{\it C.C. 67, Suc. 28, 1428 Buenos Aires, Argentina and}\\
{\it Departamento de F\'\i sica, Facultad de Ciencias Exactas y Naturales,}\\
{\it Universidad de Buenos Aires, 1428 Buenos Aires, Argentina }}
\title{The Gamow vectors and the Schwinger effect }
\maketitle

\begin{abstract}
We introduce a `proper time' formalism to study the instability of the
vacuum in a uniform external electric field due to particle production. This
formalism allows us to reduce a quantum field theoretical problem to a
quantum-mechanical one in a higher dimension. The instability results from
the inverted oscillator structure which appears in the Hamiltonian. We show
that the `proper time' unitary evolution splits into two semigroups. The
semigroup associated with decaying Gamov vectors is related to the Feynman
boundary conditions for the Green functions and the semigroup associated
with growing Gamov vectors is related to the Dyson boundary conditions.
\end{abstract}

\section{Introduction}

The history of the unification of relativity and quantum mechanics began
with the formulation of one-particle wave equations for irreducible
representations of the Lorentz group (Klein-Gordon, Dirac, etc.). However
this attempt was shown to be unsuccessful, due mainly to the appearance of
negative energy solutions which was originally considered as a strong
anomaly. Quantum field theory (QFT) arose as the solution of such
difficulties shedding light on many new results as, for example, particle
creation produced by classical fields. In a field theoretical language this
effect is originated from the instability of the vacuum of the matter field,
which privileges pair creation with respect to pair annihilation, due to the
non-symmetrical choice of vacuum, compatible with the idea of the Dirac sea
(all negative energy levels are filled). Such a process, which posed a
trouble to the one-particle interpretation of the theory, can nevertheless
be pictured as the possibility of tunneling from negative to positive energy
states under the influence of the potential barrier produced by the external
field \cite{landau,parentani}. This picture is extremely helpful for
clarifying physical ideas, so we begin explaining it in Sec. 2.

The dissipative behavior of tunneling processes in nonrelativistic quantum
mechanics has been extensively studied in the past years, and with this
instability can be associated Gamow vectors, which serves as a
representation of the exponential decay law \cite{bohm}. The aim of this
work is to study the intrinsic instability of the particle production
mechanism. However it is neither clear how to introduce Gamow vectors in the
standard first quantized theory nor at the level of second quantization. So
we consider an alternative point of view given by a consistent first quantized
formulation of relativistic quantum mechanics (RQM) based on a proper time
method.\footnote{%
From the clasical works of Fock \cite{fock}, Nambu \cite{nam}, Feynman 
\cite{fey51}, and Schwinger \cite{schwin} the proper time formalism was used in
the past for computing the effective action and studying the problem of
particle creation in external fields. In connection with this work see for
example Refs. \cite{hartle,rumpf,stephens,parentani}.} The key idea of this
formulation is to replace the Dirac solution to the negative energy problem
by the Stueckelberg interpretation \cite{stu,fey48,fey49}, i.e. to consider
(negative energy) particles travelling backward in time as antiparticles,
introducing a fifth invariant parameter for labelling the evolution of the
system. This formalism has the advantage that one can deal with a
`one-particle' configuration space of higher dimension instead of the
infinite-dimensional problem of QFT. We discuss the main outlines of this
formalism for charged scalar particles in Sec. 3, after that we consider
the problem of these particles in an external constant and homogeneous
electric field (Sec. 4) and then we introduce rigged Hilbert spaces
associated to the choice of the boundary conditions for the propagators
(Feynman and Dyson) and the corresponding proper time evolution of Gamow
vectors generated by a semigroup of unitary operators (Sec. 5), closely
related to a subjacent upsidedown oscillator structure.

\section{The physical picture}

Since the works of Heisenberg, Euler, and Kockel \cite{euler}, we know that
the vacuum is not an inert object. In fact it behaves as a dielectric in the
presence of an external electric field, and such effect introduces the
non-linear corrections to the Maxwell equations, to be discussed in Sec. 4.
Vacuum polarization corresponds to a virtual pair creation and annihilation
process, however actual pair creation can occur in the presence of external
fields. Sauter was the first in estimate the probability of creating a pair
in presence of a constant homogeneous electric field \cite{landau}. His
reasoning was made at the level of the semiclassical limit of a first
quantized theory. However they are so clear that the simple translation of
his physical picture to the intrincate theoretical framework of quantum
field theory, shed light on the understanding of the vacuum instability
problem. But before discussing the Sauter derivation, let us remember the
dictionary which allows us to translate the concepts of first quantization
to QFT.

Often, textbooks repeat that we can do relativistic quantum mechanics of the
Klein-Gordon field consistently and we can go straightforward to quantum
field theory. Certainly, this is not true. In this position wee see a
theoretical prejudice against negative energies and indefinite metric
spaces. The first obstacle was climbed up by Stueckelberg and Feynman's 
\cite{stu,fey48,fey49} and a comprehensive review of the second can be found in
the work of Feshbach and Villars \cite{feshbach}. Now let us concentrate in
the Stueckelberg and Feynman ideas using the classical picture derived from
the Klein-Gordon equation ($\hbar =c=1)$

\begin{equation}
\left( D^\mu D_\mu +m^2\right) \psi =0,  \label{kg}
\end{equation}
being $D_\mu =\partial _\mu +ieA_\mu $ the gauge covariant derivative and $%
\psi (x)$ the complex scalar field. Using the WKB approximation in (\ref{kg}%
) 
\begin{equation}
\psi (x)_{{\rm WKB}}=ae^{-iS(x)},  \label{wkb}
\end{equation}
we obtain the Hamilton-Jacobi equation in the classical limit 
\begin{equation}
(\partial _\mu S-eA_\mu )(\partial ^\mu S-eA^\mu )=m^2,  \label{hj}
\end{equation}
where $S$ is identified with the classical action and its gradient with the
canonical momentum

\begin{equation}
p_\mu =\partial _\mu S.  \label{p}
\end{equation}
The Eq. (\ref{hj}) is the mass-shell condition for spinless particles and it
is equivalent to the proper time velocity constraint ($\eta _{\mu \nu }={\rm %
diag}\{+1,-1,-1,-1\})$

\begin{equation}
\eta _{\mu \nu }\frac{dx^\mu }{ds}\frac{dx^\nu }{ds}=1,  \label{ds}
\end{equation}
due to the proportionality between the velocity and the kinematical momentum 
$\pi _\mu =p_\mu -eA_\mu $

\begin{equation}
m\frac{dx^\mu }{ds}=\pi _\mu .  \label{pi}
\end{equation}
The essence of the Stueckelberg-Feynman interpretation for antiparticles
rests on a fundamental symmetry which appears in the equations of motion of
charged particles when we reexpress them in function of the proper time.
Then we can see from Eqs. (\ref{hj}) and (\ref{pi}) that the proper time
reversal of the motion (which inverts the four-velocity, the action, and the
canonical momenta) is equivalent to a charge conjugation in these equations.
Then classical motions in which particles go backward in coordinate time ($%
\frac{dx^0}{ds}<0),$ negative kinetic energy states according to Eq. (\ref
{pi}), can be reinterpreted as charge conjugated particles (antiparticles)
going forward in time. It is important to remark that our argument can be
extrapolated to the semiclassical level. In fact, we see from the WKB wave
function that the operation which conjugates the charge in the Klein-Gordon
equation (the complex conjugation of the wave function) is equivalent to the
operation which reverses the motion in the proper time (the inversion of the
action). However a pure quantum explanation of this fundamental symmetry
requires the PFT we develop in Sec. 3, in which the concept of proper time
is introduced at the quantum level. By the moment we only use the fact that
the sign of the kinetic energy is the gauge-invariant quantity which
classifies particle and antiparticle states.

Standard QFT is nothing else than the second quantization (many-particle
theory) after changing the level of the natural vacuum (all positive kinetic
energy states unfilled) by the notion of vacuum which rest on Dirac's sea
idea for fermions (all negative kinetic energy states filled). Therefore a
pair creation in the field theoretical language (a positive-kinetic-energy
state plus a hole in the sea) corresponds to a transition from a 
negative-kinetic-energy state to a positive one in the first quantized 
theory. Note
that the charge is also negative in a negative-kinetic-energy state, in such
a way that it leaves a positive charged hole in the field theoretical
language. It turns positive when we reverse the arrow of time according to
the Stueckelberg-Feynman interpretation.

With this ideas in mind, we follow Sauter semiclassical argument. Suppose
that we have an homogeneous constant electric field ${\bf E}=E{\bf e_3.}$
Then the classical expression for a relativistic charged particle of total
energy $p_0$ is

\begin{equation}
p_0=-\sqrt{p^2+m^2}+eEx^3,  \label{E-}
\end{equation}
where we have assumed that the kinetic energy is initially negative and the
charge is $-e$ \ $(e>0)$. Suppose that we have such an incoming particle (the
outgoing antiparticle in field theoretical language) in the positive
direction of the coordinate $x^3$ \ $({\bf p=}p{\bf e_3).\ }$We see that it
decelerates in the presence of the electric field, in such a way that the
kinetic term in (\ref{E-}) decreases as the potential energy increases.
There is a turning point $a=\frac{p_0-m}{eE}$ at the coordinate $x^3$ for
this classical particle. But, quantum-mechanically, the momentum $p$ can
become imaginary ($p=ik)$ allowing the tunneling of this particle through
the barrier. Inside the barrier we have the dispersion relation

\begin{equation}
k=\sqrt{m^2-(p_0-eEx^3)^2},  \label{Eik}
\end{equation}
for the semiclassical wave associated with the particle. On the other side
of the barrier, we assume that the dispersion relation is 
\begin{equation}
p_0=+\sqrt{p^2+m^2}+eEx^3,  \label{E+}
\end{equation}
for the outgoing wave. It corresponds to a classical particle of positive
kinetic energy, which begins its movement at the point $b=\frac{p_0-m}{eE}.$

We can estimate the probability of crossing the barrier using the Gamow
formula derived from the WKB approximation

\begin{equation}
w=A\exp (-2\int_a^bk(x^3)dx^3),  \label{w}
\end{equation}
where $A$ is an undefined proportionality constant. Evaluating the integral $%
\left( t=\frac{(p_0-eEx^3)}m\right) $

\[
\int_a^bk(x^3)dx^3=-\frac{m^2}{eE}\int_{+1}^{-1}\sqrt{1-t^2}dt=\frac{\pi m^2%
}{2eE} 
\]
we finally have 
\begin{equation}
w=A\exp \left(-\frac{\pi m^2}{eE}\right).  \label{sauter}
\end{equation}
In Sec. 4 we will see that Schwinger exact calculation can be fitted by
Sauter expression choosing $A=\frac{e^2E^2}{8\pi ^3}.$ But the merit of this
heuristic derivation lies on the physical picture it draws. The vacuum of
field theory (the Dirac sea) in presence of an electric external field is
like an infinite unstable nucleus, which is disintegrated by nucleons
emission (our first quantized negative energy states going through the
barrier). Another interesting picture can be traced at classical level
solving proper time equations of motion \cite{garcia} (Lorentz force law)

\[
\frac{d^2x^\mu }{ds^2}=\frac emF^{\mu \nu }\frac{dx_\nu }{ds}. 
\]
For the case of constant electric field discussed above, adequately choosing
the initial conditions, we obtain the hyperbolic trajectories

\[
\left\{ 
\begin{array}{c}
x=\frac m{eE}\cosh (\frac{eEs}m), \\ 
\\ 
t=\frac m{eE}\sinh (\frac{eEs}m).
\end{array}
\right. 
\]

Usually, in textbooks on electrodynamics, we find only one of the branches
of the hyperbola, depending on the sign of the charge \cite{bredov}. In this
picture both particles and antiparticles evolve in the ordinary direction of
coordinate time.\footnote{%
This corresponds to the standard interpretation of QFT, which in the
canonical formulation uses the coordinate time as evolution parameter.}
However for a given charge the two solutions follows from the equations of
motion in proper time (that is analogous of what happens with the
relativistic wave equations which admit both positive and negative energy
solutions). Moreover according to the Stueckelberg and Feynman
interpretation one branch corresponds to the particle and the other to the
antiparticle. Now, let us compare the natural length-scale which appears in
the classical problem (the minimal separation between the two branches of
the hyperbola, $\frac{2m}{eE})$ with the quantum scale of the Compton
wavelength ($\frac 1m)$ given by the uncertainty principle. We see that when
these scales have the same order, particle and antiparticle trajectories
overlap increasing the probability of creating a pair. In this case the
strenght of the electric field must be of the order $E\sim \frac e{2m^2}$
according to Sauter estimation for the tunneling process.

Closing this section note that our initial vacuum is unstable under the pair
creation process and not under pair annihilation. It is due to we have
filled all negative energy states. In Sec. 3 we will see that it is
related with the analytical structure of the Feynman propagator.

\section{The proper time formalism for the charge scalar field}

In Sec. 2 we have seen that proper time can be introduced in the standard
formulation only at the semiclassical level. Such lack of an invariant
evolution parameter is the responsible of the main difficulty in reconciling
the antagonistic formalisms of quantum mechanics and the theory of
relativity \cite{gaioli}. In fact, while relativity deals with the
space-time coordinates on an equal footing, quantum mechanics privileges an
external absolute parameter to label the evolution of the state of the
system. Therefore, in the second case, `time' should have the properties of
a $c$-number, unlike in the first case where, since the spatial coordinates
are raised to the status of operators, Lorentz transformations should impose
this character on the temporal coordinate as well. Thus, this dual role of
``time'' generates a trouble in RQM. This is basically the reason why a
one-particle theory in the usual formulation of RQM finds several conceptual
difficulties.

In the forties a PTF was developed for a possible unification between
special relativity and quantum mechanics, within the framework of a
consistent one-particle theory.\footnote{%
See Refs. \cite{fanchi,ap95a,ap95b} for a review of different proposals.}
However the price payed for PFT was giving up the concept of definite mass
state. In fact, suppose that we want to develop a relativistic classical
theory, in an explicit covariant way, putting the spatial coordinates and
time in the same footing. We immediately notice that the Poisson bracket

\begin{equation}
\left\{ x^\mu ,p^\nu \right\} =-\eta ^{\mu \nu }  \label{pb}
\end{equation}
is incompatible with the classical mass constraint

\begin{equation}
p^\mu p_\mu =m^2.  \label{masa}
\end{equation}
So the classical mass constraint (\ref{masa}) must be removed from the PTF,
although nevertheless the standard RQM is recovered on-shell. Notice also
that the Poincar\'e algebra must be enlarged to contain relation (\ref{pb}).
Among the different extensions of the Poincar\'e group, whose algebra
includes the canonical commutation relation corresponding to (\ref{pb}), the
five-dimensional Galilei group \cite{roman}

\[
x^\mu \longmapsto x^\mu +v^\mu \tau , 
\]

\[
\tau \longmapsto \alpha \tau +\beta , 
\]

\[
x^\mu =L^\mu {}_\nu x^\nu +a^\mu \ ({\rm Poincar\acute e}), 
\]
is the alternative most frequently used in the literature.\footnote{%
The de Sitter group alternative is discussed in Ref. \cite{garcia}.} We
consider this formulation in this work.

Actually the parameter $\tau$, which gives the name to the formalism, is a
priori independent of the classical proper time, but it can be related to it
in the classical limit on mass shell. In the case of Galilean version of
PTF, this parameter is nothing else that the parameter of the generalized
Galilean boost, and it plays the role of a Newtonian time \cite{hor,son}. The
temporal coordinate has a different status, and it is promoted to the rank
of operator accordingly with the classical Poisson bracket (\ref{pb}). This
dissociation of roles, solves the conflict exposed above. Of course, the
notion of simultaneity and causality are the corresponding ones to Newtonian
time, but standard relativistic notions, in coordinate time, are reobtained
on shell [e.g. we will see that the retarded (causal) propagator of the
PFT, on-shell is reduced to the Feynman one].

In this way the formalism closely copies the general outlines of the
nonrelativistic quantum mechanics furnishing the theory with a well known
structure. Let us outline it for the case of the spineless relativistic
point particle.

Let $\{|x^\mu \rangle \}$ ($\mu =0,1,2,3$) be the basis of localized states
of the position operator $x^\mu $ for the charge of the system. This basis
spans a linear space endowed with a scalar product, 
\begin{equation}
\langle \Phi |\Psi \rangle =\int d^4x\Phi ^{*}(x^\mu )\Psi (x^\mu ),
\label{pe}
\end{equation}
satisfying the normalization and the completeness conditions, 
\[
\langle x^\mu |y^\mu \rangle =\delta (x^\mu -y^\mu ),\ \ \ \int d^4x|x^\mu
\rangle \langle x^\mu |=I. 
\]

In this coordinate representation the state of the system is represented by
the wave function, belonging to a four-dimensional Hilbert space, defined on
the space-time manifold. The position operator $x^{\mu \ }$and its
canonical conjugate variable, the momentum $p_\mu $, which satisfy

\begin{equation}
\lbrack x^\mu ,p^\nu ]=-i\eta ^{\mu \nu },\ \ \   \label{canonica}
\end{equation}
are given by

\[
\left\langle x\right| p_\mu \left| \Psi \right\rangle =i\partial _\mu \Psi
(x), 
\]

\[
\left\langle x\right| x^\mu \left| \Psi \right\rangle =x^\mu \Psi (x). 
\]
(In Eq. (\ref{canonica}) $-i$ was chosen to preserve the sign in the
ordinary relations for the spatial part.)

In the Schroedinger picture, 
\[
|\Psi (x^\mu ,\tau )|^2 
\]
represents the probability density for the system to be at the space-time
point $x^\mu $ at ``instant'' $\tau $. The wave function evolves with a
Schroedinger equation, 
\begin{equation}
-i\frac d{d\tau }\left| \Psi (\tau )\right\rangle =H\left| \Psi (\tau
)\right\rangle ,  \label{scho}
\end{equation}
where

\[
H=\frac{\eta ^{\mu \nu }\pi _\mu \pi _\nu }{2M} 
\]
is a super-Hamiltonian and $M$ a super-mass parameter.

The time reversal operator in the Wigner sense coincides with the charge
conjugation operator

\begin{equation}
C\Psi (x,\tau )=\Psi ^{*}(x,-\tau ).  \label{kwc}
\end{equation}
This fact naturally introduces the Stueckelberg-Feynman interpretation in
the formalism. Notice that from the Heisenberg equation of motion we have

\[
\frac{dx^\mu }{d\tau }=\frac{\pi ^\mu }M, 
\]
so, as we have mentioned above, at the classical level on the mass-shell the
parameter $\tau $ is proportional to the classical proper time

\[
d\tau ^2=\left( \frac Mm\right) ^2ds^2. 
\]

From now on, we re-scale the time calling $s=\frac 1{2M}\tau ,$ using the
Schwinger notation, which is more familiar in the literature. $s$ must not
be confused with the classical proper time. Schroedinger equation now reads

\begin{equation}
i\frac{\partial \psi (x,s)}{\partial s}=D^\mu D_\mu \psi (x,s).
\label{schos}
\end{equation}
In this case the super-Hamiltonian is re-scaled to 
\[
H=\eta ^{\mu \nu }\pi _\mu \pi _\nu =\pi ^2. 
\]
The super-Hamiltonian as well as position and momentum operators are
Hermitian in the inner product (\ref{pe}).

A stationary solution of Eq. (\ref{scho}) is 
\[
\Psi (x^\mu ,s)=e^{im^2s}\psi _{m^2}(x^\mu ), 
\]
where $\psi _{m^2}(x^\mu )$ is a solution of a generalized Klein-Gordon
equation, 
\[
D^2\psi _{m^2}(x^\mu )+m^2\psi _{m^2}(x^\mu )=0, 
\]
which is reinterpreted as an {\it eigenvalue} equation (the mass eigenvalue $%
m^2$ is real, but not restricted {\it a priori} to be positive).

We can see that the theory developed is formally identical to ordinary
quantum mechanics. Then, all we have learned from this theory can be
rewritten in the PTF. For example, let us consider the resolvent of the free
Hamiltonian operator

\[
R(z)=\frac 1{H-z}=\frac 1{p^2-z}, 
\]
We see that $R(z)$ is analytic in all the complex plane except for a cut
along the positive real axis. $R(z)$ is the extension to the complex plane
of the inhomogeneous Green function of the Klein-Gordon equation

\[
(p^2-m^2)G(m^2)=1. 
\]
The limiting values of $R(z)$ approaching the positive real axis define two
analytic functions in the lower (upper) half-plane. These functions are
those defined in QFT by adding a negative (positive) small imaginary part to 
$m^2$ in order to give sense to the formal expression 
\[
G_{\pm }(m^2)=\frac 1{p^2-m^2\pm i\epsilon },\ (\epsilon >0). 
\]
In our notation, plus and minus correspond to the Feynman and Dyson propagators
respectively 
\[
G_{F(D)}(x,y)=\left\langle x\right| G_{\pm }(m^2)\left| y\right\rangle . 
\]
These functions can be analytically continued across the cut to the second
Riemann-sheet on the upper (lower) complex half-plane, defining two analytic
functions for all $z\in {\bf C}.$

As it is well known, Feynman (Dyson) propagator can be obtained in the first
quantized theory as the inhomogeneous Klein-Gordon Green function which
propagates positive (negative) and negative (positive) energy states forward
(backward) and backward (forward) in time, respectively. In the language of
QFT Feynman propagator can be obtained as the mean value in the vacuum state
of the time ordered product of field operators

\[
G_F(x,y)=i\left\langle 0\right| \theta (x^0-y^0)\psi _{m^2}(x)\psi
_{m^2}^{\dagger }(y)+\theta (y^0-x^0)\psi _{m^2}^{\dagger}(y)\psi _{m^2}
(x)\left| 0\right\rangle 
\]
(we can write an analogous expression for the Dyson propagator, defining a
vacuum in which all positive energy states are filled.).

As we will see Feynman and Dyson boundary conditions also have an
interpretation in the off-shell theory. The two limiting values of $R(z)$
are connected with the retarded and advanced Green functions of the
Schroedinger equation. In fact, applying the formal identity 
\[
\frac 1{a\pm i\epsilon }=\mp i \int_{-\infty }^\infty \theta (\pm s)e^{i(a\pm
i\epsilon )s}ds, 
\]
for $a=H-m^2$ we see that 
\begin{equation}
G_{\pm} (m^2)=\frac 1{H-m^2\pm i\epsilon }=\mp i\int_{-\infty }^\infty \theta (\pm
s)e^{is(H-m^2\pm i\epsilon )}ds.  \label{last}
\end{equation}
Eq. (\ref{last}) in coordinate representation reads

\begin{equation}
G_{F(D)}(x,y)=\int_{-\infty }^\infty G_{\pm }[x(s),y(0)]e^{-s(m^2\pm
i\epsilon )}ds,  \label{green}
\end{equation}
with 
\[
G_{\pm }[x(s),y(0)]=\mp i\theta (\pm s)\left\langle x\right| e^{isH}\left|
y\right\rangle 
\]
the retarded and advanced solutions of the Schroedinger equation of the
off-shell theory. Eq. (\ref{green}) is the analogue of the relation between
time-dependent and -independent Green functions in nonrelativistic quantum
mechanics. The Fourier integral in $s$ selects a particular value of $m^2$
of the indefinite mass theory, and tell us that the Feynman (Dyson)
propagator is the time-independent Green function corresponding to the
retarded (advanced) one, in the off-shell theory.

Summarizing we have seen that: a) the analytical continuation of the
resolvent to the upper complex half-plane in the second sheet, b) the
boundary conditions of the on-shell Green function according to the
Stueckelberg interpretation, c) the choice of the vacuum according to the
Dirac sea idea, and d) the causal (retarded) boundary conditions of the
Green function of the off-shell theory, are the different aspects of the
same thing. In the next sections we show that in the case of the external
field problem one of the two possible analytical continuations of the
resolvent (the one associated to the Feynmam boundary conditions)
corresponds to decaying Gamov vectors in proper time, which are related to
the instability of the vacuum of field theory under pair creation.

\section{The Heisenberg-Euler effective action and the particle creation}

The effective action $S_{{\rm eff}}$ due to the interaction of the vacuum
current with the external field such that 
\[
\delta S_{{\rm eff}}=\int dx^4\left\langle 0\right| J^\mu \left|
0\right\rangle \delta A_\mu , 
\]
leads to the Heisenberg-Euler corrections of the Maxwell equations. As
it was shown by Schwinger \cite{schwin} it can be obtained in the proper
time formalism computing

\[
S_{{\rm eff}}=\int L_{{\rm eff}}(x)dx^4, 
\]
where 
\begin{equation}
L_{{\rm eff}}(x)=-i\int_0^\infty \frac{ds}s\left\langle x\right|
e^{iHs}\left| x\right\rangle e^{-i(m^2-i\epsilon )s}+C(x),  \label{leff}
\end{equation}
is the effective Lagrangian and $C(x)$ is an additive constant determined in
such a way that $L_{{\rm eff}}(x)=0$ in the absence of external fields. We
see from Eq. (\ref{leff}) that we have again reduced a field theoretical
problem (the calculation of the effective action) to a quantum-mechanical
one in a higher dimension. In fact, now our problem consists on evaluating
the matrix element $\left\langle x\right| e^{iHs}\left| x\right\rangle .$
Moreover we can reinterpret it as the persistence amplitude of the off-shell
particle to remain at the point $x^\mu $ of the space-time, and $L_{{\rm eff}%
}(x)$ as the on-shell correlate of this amplitude per unit `proper time'.

The matrix element $\left\langle x\right| e^{iHs}\left| x\right\rangle $ can
be evaluated by path integrals or in the Heisenberg picture of the canonical
formulation by means of an ingenious procedure developed by Schwinger
departing from the integration of the Heisenberg equations of motion. Here
we solve the eingenvalue problem in the Schroedinger picture. The reason is
that through this procedure it is clearer that the proper time evolution
splits into two semigroups.

Let us consider again the problem of a homogeneous and constant electric
field. Using the conventions of Sec. 3 the super-Hamiltonian reads 
\begin{equation}
H=\pi ^2=(p_0+eEx^3)^2-p_3^2-p_1^2-p_2^2.  \label{Hel}
\end{equation}
We see that it can be split in the Hamiltonian of an upsidedown harmonic
oscillator plus the Hamiltonian of two free particles in the coordinates of
the plane perpendicular to the electric field

\[
H=-H_{{\rm osc}}-p_1^2-p_2^2, 
\]

\begin{equation}
H_{{\rm osc}}=p_3+(ieE)^2(x^3+p_0/eE).  \label{hantio}
\end{equation}
The eigenvalue problem for the free particle part has the standard
plane-wave solution $\left\langle x^1x^2|p_1p_2\right\rangle .$ The inverted
harmonic oscillator has pure imaginary frequency $\frac \omega 2=\pm ieE.$
In Sec. 5 we see in detail that the two signs correspond to the spliting of
the proper time evolution of this unstable system towards the future and
towards the past respectively. The positive imaginary solution of the
eingenvalue problem, 
\begin{equation}
\left\langle x^0x^3\right| H_{{\rm osc}}\left| p_0n\right\rangle
=ieE(n+\frac 12)\left\langle x^0x^3\right. \left| p_0n\right\rangle ,
\label{vgar}
\end{equation}
corresponds to the generalized eingenstate

\begin{equation}
\left\langle x^0x^3\right. \left| p_0n\right\rangle =e^{ip_0x^0}\varphi
_n(x^3+p_0/eE),  \label{rcvga}
\end{equation}
representing a decaying Gamow vector

\begin{equation}
\left| p_0n(s)\right\rangle =e^{-eE(n+1/2)s}\left| p_0n\right\rangle ,
\label{vga}
\end{equation}
for $s>0$.\footnote{%
The positive frequency choice corresponds to positive poles of $R(z)$ in the
second sheet of the upper half-plane. This analytical extension of the
resolvent operator corresponds to the retarded (causal) evolution propagator
in proper time. Then decaying Gamov vectors are only defined for positive
times.}

Using mass `eingenstates' $\left| p_1p_2,p_0n\right\rangle $ of $H$ we can
easily evaluate the matrix element (see Appendix)

\[
\left\langle x\right| e^{iHs}\left| x\right\rangle =-\frac i{(4\pi
s)^2}\left[ \frac{eEs}{\sinh (eEs)}\right] . 
\]
Expanding it in power series of the coupling constant 
\[
\left\langle x\right| e^{iHs}\left| x\right\rangle =-\frac i{(4\pi
s)^2}\left[ 1-\frac{(eEs)^2}2+\frac 7{360}(eEs)^4+...\right] , 
\]
and replacing the third term in the integral, we finally have 
\[
L_{{\rm eff}}(x)=...-\frac 7{360}\frac{e^4E^4}{(4\pi )^2}\int_0^\infty
se^{-i(m^2-i\epsilon )s}ds=...+\frac 7{360}\frac{e^4E^4}{m^4}, 
\]
which coincides with the expression obtained by Schwinger \cite{schwin} for
the Heisenberg-Euler correction in the spinless case.

Now let us discuss the particle creation process associated with the
imaginary part of

\[
L_{{\rm eff}}(x)=-\frac 1{(4\pi )^2}\int_0^\infty \frac{eEs}{s^3\sinh (eEs)}%
e^{-i(m^2-i\epsilon )s}ds+C(x). 
\]
We can analytically continue the integrand to the lower half-plane, then the
integral along the positive real axis becomes

\[
L_{{\rm eff}}(z)=-\frac 1{(4\pi )^2}\int_{\Gamma _0}\frac{eE}{z^2\sinh (eEz)%
}e^{-im^2z}dz+C(x), 
\]
where $\Gamma _0$ is a path with the same end points. The integrand has
poles in $z_{\pm n}=\pm i\frac{n\pi }{eE}.$\footnote{%
Note also that the resolvelt 
\[
R(z)=\frac 1{\pi ^2-z}, 
\]
has a cut on the positive real axis and poles $z_n=\pm ieE(n+\frac 12)$ in
the positive imaginary one. Since the effective Lagrangian can be written as 
$L_{{\rm eff}\ }(z)$ =$-i\left\langle x\right| \ln R(z)\left|
x\right\rangle +C(x)$, the poles of the resolvent are related with the poles
of the integrand.} Using the residues theorem we can rewrite $L_{{\rm eff}%
}(z) $ as 
\[
L_{{\rm eff}}(z)=-\frac 1{(4\pi )^2}\int_{\Gamma _{-1}}\frac{eE}{z^2\sinh
(eEz)}e^{-im^2z}dz+C(x)-2\pi i{\rm Res}(z_{-1}), 
\]
where the path $\Gamma _{-1}$ has the same end points as $\Gamma _0$ in such
a way that the closed counterclockwise contour $\Gamma _{-1}\cup (-\Gamma
_0) $ encircles the first pole in the negative imaginary axis, $%
z_{-1}=-i\frac \pi {eE}.$ The probability of creating one-pair\footnote{%
The contribution corresponding to create two-pairs can be obtained taking a
path which encircles the pole $z_{-2},$ and so on.} per space-time volume is
given by

\[
w(x)={\rm Im}L_{{\rm eff}}(x)=-2\pi {\rm Res}(z_{-1}). 
\]
Thus we finally have

\begin{equation}
w(x)=\frac{(eE)^2}{8\pi ^3}\exp \left( -\frac{\pi m^2}{eE}\right) ,
\label{prounp}
\end{equation}
according to Sauter's estimation.

\section{Gamow vectors associated with particle creation by an external field}

In Sec. 4 we have seen that an inverted harmonic oscillator structure
appears in the Hamiltonian. 
We claim that the instability of the vacuum in the
field theoretical language has its correlate in this simple unstable system
at the level of the off-shell theory. 
We have noticed that this inverted
oscillator has a pure imaginary frequency $\omega =\pm ieE$
(antioscillator). This fact leads to a `complex eigenvalue' problem which 
requires some technical points we are going to discuss in this section.

We can solve the eingenvalue problem for this system from the solutions of
the harmonic oscillator, obtaining two sets of `complex eigenvalues' (which
do not take part of the spectrum of $H_{{\rm osc}})$ 
\begin{equation}
\left\langle x^0x^3\right| H_{{\rm osc}}\left| p_0n\right\rangle
=ieE(n+\frac 12)\left\langle x^0x^3\right. \left| p_0n\right\rangle ,
\label{gvmas}
\end{equation}

\begin{equation}
\left\langle x^0x^3\right| H_{{\rm osc}}\left| \widetilde{p_0n}\right\rangle
=-ieE(n+\frac 12)\left\langle x^0x^3\right. \left| \widetilde{p_0n}%
\right\rangle ,  \label{gvmen}
\end{equation}
being

\begin{equation}
\left\langle x^0x^3\right. \left| p_0n\right\rangle =e^{ip_0x^0}\varphi
_n(x^3+p_0/eE),  \label{hermas}
\end{equation}

\begin{equation}
\left\langle x^0x^3\right. \left| \widetilde{p_0n}\right\rangle =e^{ip_0x^0}%
\widetilde{\varphi _n}(x^3+p_0/eE),  \label{hermen}
\end{equation}
where $\varphi _n$ and $\widetilde{\varphi _n}$ contain Hermite polynomials
of complex argument. The `eigenvalues' correspond to complex poles $z_n=\pm
ieE(n+1/2)$ along the positive/negative imaginary axis of the resolvent
operator $R(z)$. It was demonstrated \cite{diener} that the corresponding
`eigenvectors' form a biorthogonal set, i.e.

\begin{equation}
\left\langle \widetilde{p_0n}\right. \left| p_{0^{\prime }}m\right\rangle
=\delta \left( p_0-p_{0^{\prime }}\right) \delta _{nm},\hspace{0.3in}{\rm in}%
\ \Phi _{+},  \label{orto}
\end{equation}

\begin{equation}
\left\langle p_0n\right. \left| \widetilde{p_{0^{\prime }}m}\right\rangle
=\delta \left( p_0-p_{0^{\prime }}\right) \delta _{nm},\hspace{0.3in}{\rm in}%
\ \Phi _{-},  \label{orton}
\end{equation}

\begin{equation}
\int dp_0\sum\limits_{n=0}^\infty \left| p_0n\right\rangle \left\langle 
\widetilde{p_0n}\right| =I,\hspace{0.3in}{\rm in}\ \Phi _{+},
\label{compf}
\end{equation}

\begin{equation}
\int dp_0\sum\limits_{n=0}^\infty \left| \widetilde{p_0n}\right\rangle
\left\langle p_0n\right| =I,\hspace{0.3in}{\rm in}\ \Phi _{-}.
\label{compd}
\end{equation}
The eigenvectors (\ref{hermas}) and (\ref{hermen}) correspond to generalized
eigenvectors of $H_{{\rm osc}}$ in adequate rigged Hilbert spaces:

\begin{equation}
\left\langle H_{{\rm osc}}\phi \right. \left| p_0n\right\rangle
=\left\langle \phi \right| H_{{\rm osc}}\left| p_0n\right\rangle =ieE\left(
n+\frac 12\right) \left\langle \phi \right. \left| p_0n\right\rangle ,
\label{gemas}
\end{equation}

\begin{equation}
\left\langle H_{{\rm osc}}\psi \right. \left| \widetilde{p_0n}\right\rangle
=\left\langle \psi \right| H_{{\rm osc}}\left| \widetilde{p_0n}\right\rangle
=-ieE\left( n+\frac 12\right) \left\langle \psi \right. \left| \widetilde{%
p_0n}\right\rangle ,  \label{gemen}
\end{equation}
where $\phi \in \Phi _{+}$ and $\psi \in \Phi _{-},$ since $H_{{\rm osc}%
}^{\dagger }=H_{{\rm osc}}$ is continuous on $\Phi _{\pm }$. The test spaces 
$\Phi _{+}$ and $\Phi _{-}$ are defined by \cite{diener}

\begin{equation}
\Phi _{+}=\left\{ \phi \in {\cal S}/\left\langle v\right. \left| \phi
\right\rangle \in {\cal Z}\right\} =\left\{ \phi \in {\cal S}/\left\langle
u\right. \left| \phi \right\rangle \in {\cal K}\right\} ,  \label{fimas}
\end{equation}

\begin{equation}
\Phi _{-}=\left\{ \psi \in {\cal S}/\left\langle v\right. \left| \psi
\right\rangle \in {\cal K}\right\} =\left\{ \psi \in {\cal S}/\left\langle
u\right. \left| \psi \right\rangle \in {\cal Z}\right\} ,  \label{fimen}
\end{equation}
where ${\cal K}$ is the subset of Schwarz functions (${\cal S}$) of compact
support and ${\cal Z}$ is the subset of Schwarz integer functions of
exponential order, restricted to the real axis. \{$\left| v\right\rangle $\}
and \{$\left| u\right\rangle $\} are two representations constructed with
the generalized eigenvectors of the operators

\begin{equation}
v=\frac 1{\sqrt{2}}\left[ \frac{p_3}{\sqrt{eE}}+\sqrt{eE}\left( x^3+\frac{p_0%
}{eE}\right) \right] ,  \label{v}
\end{equation}

\begin{equation}
u=\frac 1{\sqrt{2}}\left[ \frac{p_3}{\sqrt{eE}}-\sqrt{eE}\left( x^3+\frac{p_0%
}{eE}\right) \right] ,  \label{u}
\end{equation}
which are the analogues of the creation and annihilation operators for the
harmonic oscillator. Therefore we can construct a pair of rigged Hilbert
spaces:

\begin{equation}
\Phi _{\pm }\subset {\cal L}^2({\bf R})\subset \Phi _{\pm }^{\times },
\label{rhs}
\end{equation}
where ${\cal L}^2({\bf R})$ is a Hilbert space, $\Phi _{\pm }$ are dense
subsets of ${\cal L}^2({\bf R})$ with their own complete nuclear topology,
and $\Phi _{\pm }^{\times }$ are the dual spaces of $\Phi _{\pm }$. The
evolution operator $U=e^{iHs}$ is continuous on $\Phi _{\pm }$ and such that 
$U^{\dagger }\Phi _{\pm }\subset \Phi _{\pm },$ for $s {>\atop <} 0$
only$.$ Thus we can obtain the evolution in $s$ of the pair of Gamow vectors
as

\begin{equation}
\left| p_0n(s)\right\rangle =e^{-eE(n+1/2)s}\left| p_0n\right\rangle ,{\rm 
\qquad for }s>0,  \label{smas}
\end{equation}

\begin{equation}
\left| \widetilde{p_0n(s)}\right\rangle =e^{eE(n+1/2)s}\left| \widetilde{p_0n%
}\right\rangle ,{\rm \qquad for }s<0,  \label{smen}
\end{equation}
which are functionals in $\Phi _{\pm }^{\times },$ respectively. We see that
the pair of Gamow vectors represent a decaying state towards the future and
a growing state from the past, respectively. These Gamow vectors are related
via the $s$-time reversal Wigner operator $K,\footnote{%
The Wigner operator $K$ is such that conjugates the wave function of the
Klein-Gordon equation and coincides with the charge conjugation operator for
this equation. The $s$-reversal operation of Eq. (\ref{schos}) is given by
the operator $S$ such that $S\Psi (x,s)=K\Psi (x,-s),$ which coincides with
the generalization of the charge conjugation operation of Eq. (\ref{schos}).}
$ since it can be proved \cite{diener} that $K:\Phi _{\pm }\longrightarrow
\Phi _{\mp },$ and therefore

\begin{equation}
K:\Phi _{\pm }^{\times }\longrightarrow \Phi _{\mp }^{\times }.  \label{kwig}
\end{equation}

We have seen that the unitary temporal evolution splits into two semigroups.
Let us interpret the physical meaning of such splitting. Coming back to the
off-shell propagator let us evaluate it for the Hamiltonian (\ref{Hel}):

\begin{eqnarray}
G_{+}(x,y,s) &=&\theta (s)\left\langle x\right| e^{iHs}\left| y\right\rangle
\label{fpdp} \\
\ &=&\theta (s)\int dp_1\int dp_2\int dp_0\sum\limits_{n=0}^\infty \left\langle
x\right| e^{iHs}\left| p_1p_2p_0n\right\rangle \left\langle \widetilde{%
p_1p_2p_0n}\right| \left. y\right\rangle ,  \nonumber
\end{eqnarray}
where we have used the completeness relation (\ref{compf}) since their
generalized eigenvectors are well defined only for $s>0.$ It shows that the
retarded condition in time $s$ (Feynman prescription) is satisfied only for
Gamow vectors decaying towards the future$.$ Similarly the Dyson
prescription is related to growing Gamow vectors.

We can conclude that the choice of Feynman or Dyson prescriptions is {\it a
priori }a conventional matter when we consider the whole Universe (or a
closed isolated system). That is, for example, Feynman boundary condition
for which particles propagate towards the future and antiparticles towards
the past in coordinate time, is correlated with the Gamow vectors
decaying towards the future in proper time. Then we have a unavoidable
proper time asymmetry, coming from a proper time symmetrical theory which splits
the dynamical evolution into two semigroups. But once we have made 
the Feynman 
choice, `proper time' asymmetry is a substantial thing providing a privileged
direction of time, the one in which Gamow vectors decay. Dyson choice only
leads to a specular world, in which we can be living just now, if we
conventionally interchange the role of past and
future. 

\section*{Appendix}

The matrix element $\left\langle x\right| e^{iHs}\left| x\right\rangle $ can
be factorized as the matrix element for two free particles and a upside down
harmonic oscillator 
\[
\left\langle x^1\right| e^{-i(p_1)^2s}\left| x^1\right\rangle \left\langle
x^2\right| e^{-i(p_2)^2s}\left| x^2\right\rangle \left\langle x^0x^3\right|
e^{i(H_{{\rm osc}})s}\left| x^0x^3\right\rangle . 
\]
The first factor gives a contribution 
\begin{eqnarray*}
\left\langle x^1\right| e^{-i(p_1)^2s}\left| x^1\right\rangle
&=&\int_{-\infty }^\infty \left\langle x^1|p_1\right\rangle \left\langle
p_1\right| e^{-i(p_1)^2s}\left| x^1\right\rangle dp_1 \\
&& \\
\ &=&\frac 1{2\pi }\int_{-\infty }^\infty e^{-i(p_1)^2s}dp_1=\frac 1{2\pi }%
\sqrt{\frac \pi {is}},
\end{eqnarray*}
and the same factor for the second one. The third factor can be computed in
a similar way 
\[
\left\langle x^0x^3\right| e^{i(H_{{\rm osc}})s}\left| x^0x^3\right\rangle =%
\sum_n\int_{-\infty }^\infty \left\langle x^0x^3\right|
e^{i(H_{{\rm osc}})s}\left| p_0n\right\rangle \left\langle \widetilde{p_0n}
\right| \left. x^0x^3\right\rangle dp_0 
\]
\[
=\sum_n\int_{-\infty }^\infty e^{-eE(n+\frac 12)s}
\varphi _n (x^3+p_0/eE)\widetilde{\varphi _n}(x^3+p_0/eE)dp_0 
\]
\[
=\frac{eE}{2\pi }\sum_n e^{-eE(n+\frac 12)s}\int_{-\infty
}^\infty \widetilde{\varphi _n}(u)\varphi _n(u)du=\frac{eE}{2\pi }\left[
\frac 1{2\sinh (eEs)}\right] . 
\]
Collecting our partial result we finally have 
\[
\left\langle x\right| e^{iHs}\left| x\right\rangle =-\frac i{(4\pi
s)^2}\left[ \frac{eEs}{\sinh (eEs)}\right] . 
\]

\section*{Acknowlegments}

The authors are grateful to the organizers of the First International
Colloquium on `Actual Problems in Quantum Mechanics, Cosmology, and the
Primordial Universe' and the `Foyer d'Humanisme' for their warm hospitality
in Peyresq.

This work was partially supported by grants C1I$^{*}$-CI94-0004 of the
European Community, PID-0150 of CONICET, 
EX-198 of Universidad de Buenos, and 12217/1 of Fundaci\'on
Antorchas and British Council.

\end{document}